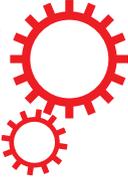

# Optical vortex knots – one photon at a time

Sebastien J. Tempone-Wiltshire, Shaun P. Johnstone & Kristian Helmerson

Feynman described the double slit experiment as "a phenomenon which is impossible, *absolutely* impossible, to explain in any classical way and which has in it the heart of quantum mechanics". The double-slit experiment, performed one photon at a time, dramatically demonstrates the particle-wave duality of quantum objects by generating a fringe pattern corresponding to the interference of light (a wave phenomenon) from two slits, even when there is only one photon (a particle) at a time passing through the apparatus. The particle-wave duality of light should also apply to complex three dimensional optical fields formed by multi-path interference, however, this has not been demonstrated. Here we observe particle-wave duality of a three dimensional field by generating a trefoil optical vortex knot – one photon at a time. This result demonstrates a fundamental physical principle, that particle-wave duality implies interference in both space (between spatially distinct modes) and time (through the complex evolution of the superposition of modes), and has implications for topologically entangled single photon states, orbital angular momentum multiplexing and topological quantum computing.

Since Einstein's description of the photon in 1905[1], used to explain the photoelectric effect, the quantized nature of light has been accepted. Numerous experiments since have shown the particle-wave duality of photons and other quantum objects. These include two slit interference experiments with single photons[2–4], electrons[5–7], atoms[8] and even much larger macroscopic objects such as $C_{60}$ buckyballs[9]. In the realm of single photons, more complex intensity and phase structures have been investigated including both analog[10] and digital[11,12] holograms at ultralow light intensities. It has also been shown that single photons can carry the orbital angular momentum associated with optical vortices[13]. In all of these experiments, the intensity (and phase) profiles resulting from interference were essentially two-dimensional (2D). Further propagation in the far-field produced a self-similar 2D pattern.

Here we report the first measurements of a three dimensional (3D) complex optical field - one photon at a time - by demonstrating that the distribution of single photons evolves to the full complex field. These results are unique in that not only do we measure a complex three dimensional vortex structure within the optical field, produced one photon at a time, but we also show that the effects of diffraction on a single photon are the same as for the entire optical field, even though diffraction is a 'non-local' effect.

The three dimensional knotted structure investigated in this paper is made from optical vortices. An optical vortex[14] is a point in an optical field about which the field undergoes some integer multiple of $2\pi$ phase winding, leading to an intensity zero at the vortex core as at this point the phase of the field is undefined. These vortices have been shown to carry the orbital angular momentum of an optical field and thus can be both 'positive' and 'negative', depending upon the sense of rotation of the angular momentum. They therefore arise from the global structure of the field, not the local structure. These vortices trace out lines along the direction of propagation of the optical field that can only be removed by a transfer of angular momentum to some object they interact with, or through annihilation with a vortex of opposite sign.

An optical vortex knot is a vortex line in an optical field that follows the multiply-connected topology of a knot. As such, it is an inherently three-dimensional optical field that exhibits both complex, but measurable, phase and intensity profiles. Optical vortex knots can be constructed in the paraxial approximation by a superposition of Laguerre-Gaussian (LG) beams[15]. Isolated optical vortex knots, including the trefoil (Fig. 1b), cinquefoil and a Hopf link were demonstrated in the laboratory[15] by generating the appropriate superposition of LG modes using a spatial light modulator (SLM).

We generate our optical field employing methods similar to those of refs 15 and 16, using a complex valued, phase only hologram[17] displayed upon a SLM. A reference Gaussian beam is also generated from the SLM and made to interfere with the knotted optical vortex field. The vortices leave a characteristic forked structure in the

School of Physics and Astronomy, Monash University, Victoria 3800, Australia. Correspondence and requests for materials should be addressed to S.J.T.-W. (email: sebastien.tempone-wiltshire@monash.edu)

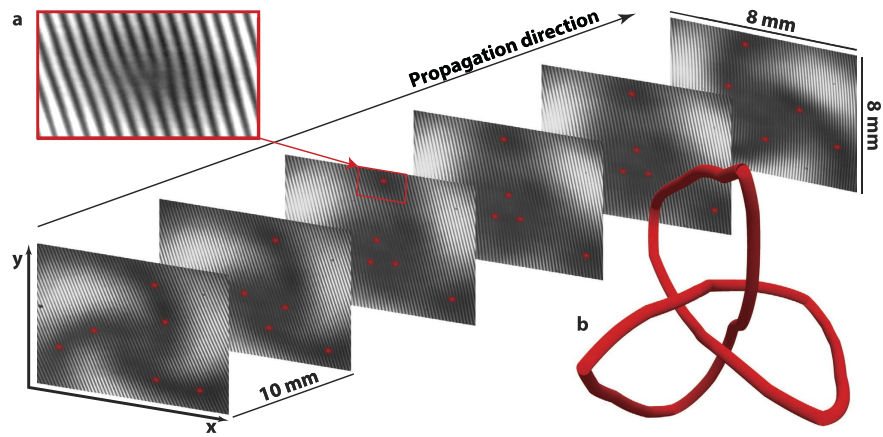

**Figure 1. Measuring the optical fields knotted topology.** Different transverse planes of the optical field along the beam propagation direction (indicated by the black arrow) are imaged onto the camera by moving the position of the final imaging lens before the camera (see optical set up in left of Fig. 2). The length scales in both the propagation direction and transverse plane are shown. (**a**) Vortices within each plane (indicated by red dots) are located by the characteristic forked structure they create in each interferogram, due to the $2\pi$ phase winding about each vortex. (**b**) The vortices located in 25 separate planes are then 'stitched' together to form the three dimensional trefoil knot nodal structure.

resulting interferogram, as shown in Fig. 1a, which facilitates identifying their locations. A schematic of the optical set up is shown on the left in Fig. 2, with the inset showing the hologram used to generate the knotted vortex structure and the reference Gaussian beam. A number of different transverse planes of the knotted optical field (Fig. 1) are imaged onto the camera by translating a final imaging lens (which also provides magnification of the field) in the direction of propagation. The vortices within each image plane are located and then 'stitched' together to visualise the three dimensional trefoil knot vortex structure (Fig. 1b).

To demonstrate that the optical vortex knot can be generated with single photons, we perform the measurement in a regime where on average there is less than a single photon in the apparatus at any given time. The SLM is illuminated with a laser intensity of $<10\,\mathrm{pW}$ corresponding to an average of $<0.1$ photons along the $\approx 1$ metre path from the SLM to the detection plane. Assuming the laser emits a Poissonian distribution of photons in time[18,19], the likelihood of observing two photons along this path is 20 times less than for a single photon. Additionally, due to losses from other optical elements and diffraction from the SLM into other optical modes, the average number of photons along the optical path is further ($\approx 20$ times) reduced, placing the optical field well within the single photon regime.

We detect the optical field at the single photon level using an electron multiplying CCD (EMCCD) camera with high gain in photon counting mode, where each pixel can at most register detection of a single event. Exposure times were on the order of $0.3$ seconds corresponding to the detection of $\approx 2500$ photons in each image, which is much fewer than the number of camera pixels, ensuring only single photons would be detected per pixel. Figure 2a (centre) shows a single acquired image, where each pixel is either assigned a value of 1 or 0 upon read-out as either a photon has entered a pixel within the exposure time, or not. To build up high enough contrast to locate the vortices, a series of images is taken, on the order of 8,000 images, and summed together to form a final image representing the intensity of the optical field within a given plane. The result of the summation of 100, 1000 and 8000 images is shown in Fig. 2b–d (centre), respectively. (Supplementary Video 1 shows the summation of increasing numbers of images at a single plane.)

It is clear that the greater number of images taken, the greater the contrast in the measurement, however, this is limited in practice by two factors. Firstly, the dark noise cannot be completely removed and creates a 'noise floor' which decreases the contrast with increasing number of images. The second limiting factor is the interferometric stability of the set up, as small drifts in path length difference, on the order of $260\,\mathrm{nm}$, can completely wash out the fringes. Thus all data must be taken over the time the fringes are stable, on the order of $30\,\mathrm{minutes}$.

To verify the knotted topology of the optical field from the detection of single photons, we took a series of 30 images in different transverse planes along the propagation direction of the knot, some of which are shown in Fig. 1. Each of these images consisted of the summation of 8000 individual measurements, as described above, to allow the vortices to be accurately located. The locations of these vortices were then 'stitched' together to generate the resulting knotted nodal structure shown in Fig. 1b. (Supplementary Video 2 shows a rotating view of the knotted 3D topology.) This shows that even measured at the single photon level, the resultant optical field has the knotted topology.

As a quantitative measure of how each single image contributes to the equivalence between the measured and the predicted knotted optical field, we performed a normalised cross correlation (NCC) of the data at $Z=0$ (the waist of the beam), as each new image was added. Figure 2 (right) shows the peak value of the computed NCC as a function of photon number. For comparison we also computed the NCC of the data with a theoretical Gaussian beam and a modal decomposition of the knot. Each of the curves seem to follow a similar shape, initially growing approximately linearly, followed by a rapid flattening of the curve once the total photon number approaches $\approx 10^6$ photons, corresponding to on average $\approx 3$ photons per pixel. Interestingly, even from a single image containing only $\approx 600$ photons, corresponding to on average only 0.002 photons per pixel, there is already a considerable difference in the correlation values, with values of 0.08, 0.06 and 0.04 for the correlation with the knot, the Gaussian and the modal decomposition respectively. These curves show that relatively few photons, $\approx 10^5$, are required to be able to adequately distinguish the resultant optical field from the comparison fields.

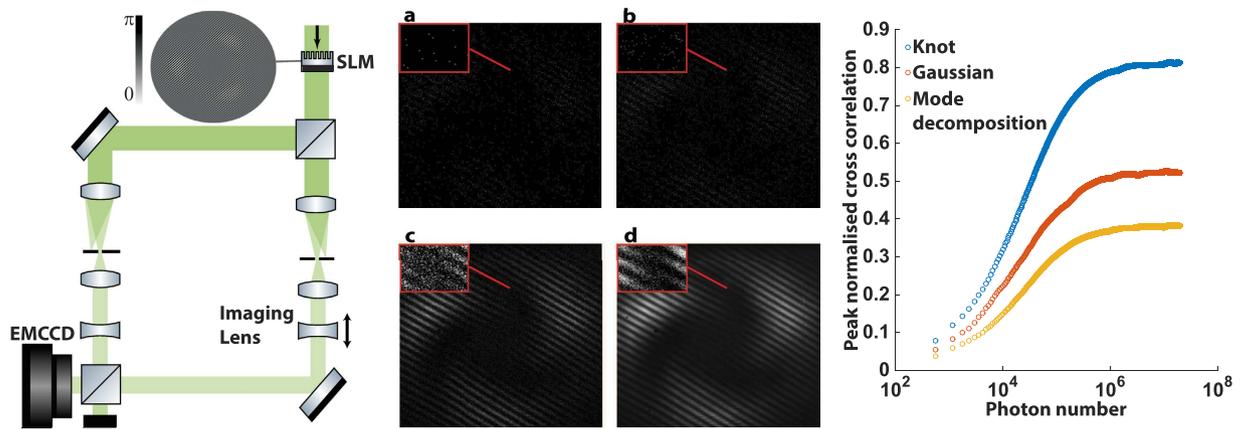

**Figure 2. Left: Optical set-up.** Each photon passes through the spatial light modulator (SLM), which acts as a static phase element generating both a knot and a Gaussian reference beam, which is split and then spatially filtered such that one path contains the knot, and the other the reference beam. These two paths are then recombined at a small angle, which generates an interferogram that facilitates identification of the vortex structure, and then imaged onto the camera. The double headed arrow indicates the lens that is translated to image different transverse planes of the knot. The inset shows the hologram displayed on the SLM (available as supplementary material). Centre: Interferometric measurements of the knotted topology. Measurements of the optical field are made by summing multiple images that each contain an average of much less than one photon per pixel. (**a**) A single frame imaged by the camera in photon counting mode. (**b–d**) Measurements resulting from the summation of 100, 1000 and 8000 images, respectively, with the contrast of the resultant interference pattern increasing with increasing numbers of images making the forked vortex structure more visible (shown magnified in insets). Right: Normalised cross correlation (NCC) as a function of photon number. The peak value of the NCC between the interferometric measurement of the knotted optical field at $Z = 0$ with the expected theoretical intensity pattern of the knot, a Gaussian mode and the modal decomposition of the knot used in the 'which path' measurement, each interfered with a Gaussian mode to reproduce the global fringe pattern as a function of increasing number of summed images, displayed as increasing photon number.

We further verified that the distribution of single photons produces the knotted nodal structure via a 'which path' measurement. The optical set up used for this measurement, shown on the left in Fig. 3, involves splitting the optical path at the SLM so that each path only passes through one half of the SLM screen. This path separation allows a chopper wheel to be inserted just after the SLM, which at any given time blocks one of the two paths.

The paraxial approximation of a trefoil optical vortex knot can be decomposed into 5 different Laguerre–Gaussian modes with indices $(\rho, m) = (0, 0), (1, 0), (2, 0), (3, 0)$ and $(0, 3)$[15]. In this 'which path' measurement, a hologram for the complex field resulting from the summation of the first 4 modes is displayed on the left-hand side of the SLM, while the fifth mode is displayed on the right hand side, as indicated in the inset of Fig. 3. This division of the modes was chosen as each of the resultant fields from the holograms retains its rotational symmetry, while the knot's intensity profile does not. These two fields are then recombined to form the knotted optical field, and imaged onto the camera with a pair of lenses.

We show that knowledge of the photons trajectory destroys the knotted optical field by switching on the chopper wheel, which alternately blocks the optical path from either side of the SLM at a rate of 1 kHz. The right hand side of Fig. 3 shows three columns, the first (leftmost) is the theoretical intensity profile of the knotted optical field, the second (middle) is the resultant intensity profile of the field when the two optical modes are allowed to interfere, and the third (rightmost) is the pattern obtained with the chopper wheel is inserted. Each column consists of images at 0, 5 and 10 cm along the direction of propagation, respectively, showing the evolution of the optical field. Comparison of the first and second column clearly shows qualitatively the agreement of the intensity profile of a knotted optical field in each plane, however, to verify this we performed a NCC of the intensity data with theory and found peak values of 0.91, 0.89 and 0.89 for $Z = 0$, 5 and 10 cm, respectively.

The third column shows the result of the chopper wheel being inserted. This appears to restore the radial symmetry of the two optical fields as, since the photon cannot traverse both paths at once, the two complex fields cannot interfere with one another and thus the resultant field is simply a sum of the intensity of the two individual fields. The peak NCC of this data with the expected knotted optical field (first column) is 0.67, 0.74 and 0.46, which is significantly lower than the respective values obtained for the second column. Furthermore, NCCs of the unchopped optical field (second column) yielded an average peak value of 0.54, while NCCs between the chopped field yielded an average peak value of 0.75, indicating that the chopped optical field had a much stronger self similarity upon propagation than the knotted optical field, and therefore less three dimensional structure.

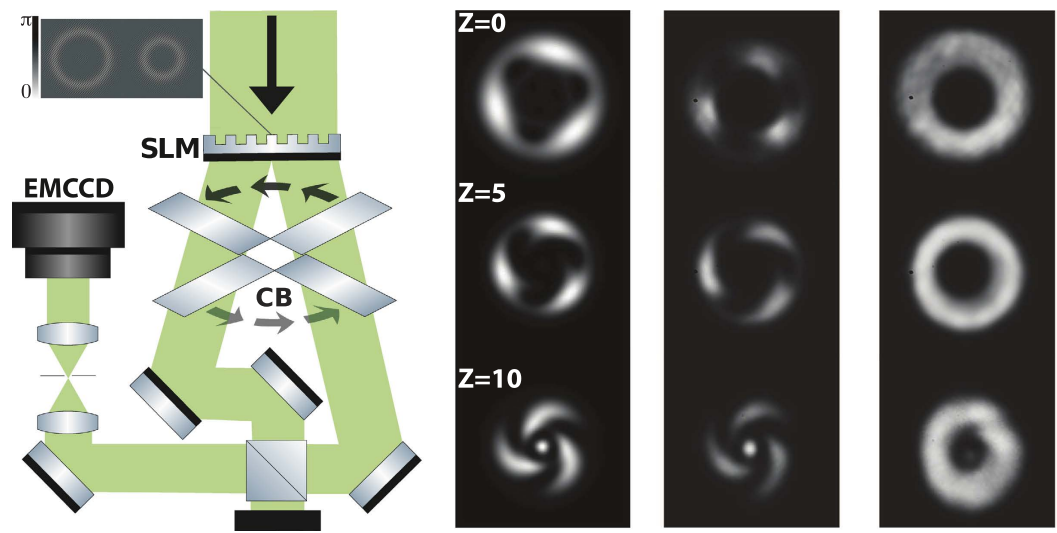

**Figure 3. Left: Optical set-up for a 'which path' measurement.** The SLM contains two holograms (shown in the inset and provided as supplementary material) side by side. The laser beam is split into two and passes through a chopper blade (CB), which ensures that photon's are present along the optical path from only one of the two holograms at any given time. The two beam paths are later recombined and imaged onto the camera. Right: 'Which path' measurements of the optical field. The three columns shown contain the theoretical intensity patterns of the knotted optical field (left), the intensity patterns resulting from the unchopped beams (centre) and from the chopped beams (right). These images are taken at three different transverse planes along the beam path (positions are indicated in centimetres). A normalised cross correlation of the data (centre and right) with simulation (left) shows a higher correlation of the knot with the unchopped optical field, with an average peak value of 0.89, as opposed to when the beam is chopped which yields an average peak value of 0.62.

In summary, we have demonstrated that a distribution of single photons can recreate the complex three dimensional structure of a superposition of optical modes. The topology of the optical field was measured by tracing the vortex structure along the direction of propagation of the optical field from the ensemble spatial measurement of single photons. This is the first measurement of a 3D complex intensity and phase structure within an optical field, imaged in the single photon regime, confirming that single photons diffract in the same manner as the entire optical field. A 'which path' measurement was also made by generating a knotted 3D topological structure through two spatially separate holograms, only allowing photons from one hologram at any time along the optical path. This measurement shows knowledge of the photon's trajectory destroys the resultant topological structure – each photon must interact with both holograms, being in a superposition of different optical modes.

## Methods

**Optical set up.** A continuous wave, frequency doubled, Nd:YAG laser emitting approximately 15 mW of power at 532 nm was used as the light source. The photons emitted from this source pass through a 532 nm filter, a polariser, a half wave plate and a number of neutral density filters before being spatially filtered to produce a vertically polarised, pure $TEM_{00}$ mode with ≈7 pW of power for the experiments. As indicated in the left hand side of Figs 2 and 3, non-polarising beam splitters were used to generate and recombine the multiple modes used within this experiment, and multiple telescopes and apertures were used for spatially filtering the optical field as well as for magnification. The spatial light modulator (SLM) is a liquid crystal device that acts as a phase only, diffractive optical element that has a maximum phase shift of $1.19\pi$ at 532 nm. The digital holograms were encoded using the methods described in ref. 17.

**Generating a knotted optical field.** To generate an optical field with a knotted vortex structure, the complex scalar field that diffracts into the knotted structure must first be known. We use a similar construction to ref. 15 to generate an isolated vortex knot with the topology of a trefoil knot, however, we introduce two new parameters which alter the geometry of the vortex structure while retaining its topology. To calculate the complex scalar field, the trajectories of the vortices are first defined within a cylinder, with the number of vortices, and the number of 'turns' the vortices perform within the length of the cylinder determining the topology of the knot. For a trefoil knot, we begin by defining two nodal lines on circular path trajectories, both of which undergo one and a half turns along the length of the cylinder but are $\pi$ out of phase. Our modification amounts to altering the geometry of this structure by generalising the trajectory of the nodal lines to ellipses instead of circles, i.e.

$$e^{i\frac{3\theta}{4}} \rightarrow \frac{a(1-b^2)e^{i\frac{3\theta}{4}}}{1+b\,\cos\left(\frac{3\theta}{4}\right)},$$

(1)

introducing two new parameters, $a$ and $b$, which determine the relative size and eccentricity of the elliptical orbit, respectively. Here we have parameterised the trajectories in terms of $\theta$, an angle from some chosen reference point. Within certain ranges of values, these two parameters allow the geometry of the knot to be altered without destroying the topology. In particular the aspect ratio of the knot, defined as the length over which the knot forms for a given beam size, can be increased or decreased by changing the parameter '$a$' as this alters the radial separation of the nodal lines. For more details, refer to the supplemental information.

**Detecting single photons.**    An Andor iXon electron multiplying charged coupled device (EMCCD) camera with high gain is used. The camera is cooled to $-100\,°C$, achieving the lowest readout and 'dark' noise possible. Photon counting mode was enabled, and the discriminator set to maximise the signal to noise ratio, with the noise defined as the total counts when the shutter was closed for the given exposure time. Exposure times were limited to a maximum duration of $\approx 0.3$ seconds as only single photons per pixel could be detected in photon counting mode. For a 0.3 second exposure, at an intensity of 7 pW one would expect $\approx 5 \times 10^6$ photons to hit the camera within an exposure, however, due to optical losses only approximately 1% of the photons incident on the SLM will make it to the detector, yielding a maximum of $5 \times 10^4$ photons hitting the EMCCD within an exposure. As the EMCCD display is 512 by 512 pixels, this gives an average 0.2 photons per pixel across the EMCCD, which is below the single photon per pixel limit.

**Locating optical vortices.**    The location of the optical vortices was found through interferometric measurements of the phase structure, with characteristic forks in the fringe pattern indicating the presence of an optical vortex. The precision to which one can identify these vortex structures is set by the fringe spacing, which is determined by the relative angle of the direction of propagation for the knotted optical field and the Gaussian reference beam. If two vortices are closer together than the fringe spacing, then instead of two individual vortices they appear as a single vortex whose phase winding is the summation of the two individual vortices. Thus to maximise how precisely the vortices could be located, the relative angle between the Gaussian reference beam and the knotted optical field was varied until the fringe spacing was on the order of a few pixels across, ensuring the fringes would still be easily visible within the image.

**Normalised cross correlations of the data.**    All normalised cross correlations of the data with the theory were performed in Matlab with the inbuilt 'normxcorr2' function. The values quoted within the text are the peak values of the resulting matrices. For the data set shown on the right-hand side of Fig. 3 exposure times were shorter, allowing $\approx 600$ photons to hit the CCD within an exposure so as to increase our resolution in the resulting measure of normalised cross correlation as a function of photon number.

## Acknowledgements
We thank T. Hughes and A. Bishop for their comments on the manuscript. S.J.T.-W. thanks R. Anderson for help stitching the vortices together in *Mathematica.* This work is supported by Australian Research Council (ARC) Discovery Project grant DP130102321.

## Author Contributions

S.J.T.-W. and S.P.J. performed the experiments. K.H. and S.J.T.-W. conceived and designed the experiments. All authors co-wrote the paper.

## Additional Information

**Supplementary information** accompanies this paper at http://www.nature.com/srep

**Competing financial interests:** The authors declare no competing financial interests.

**How to cite this article**: Tempone-Wiltshire, S. J. *et al.* Optical vortex knots – one photon at a time. *Sci. Rep.* **6**, 24463; doi: 10.1038/srep24463 (2016).